\documentstyle[sprocl,wrapfig,epsfig]{article}

\def\Journal#1#2#3#4{{#1} {\bf #2}, #3 (#4)}

\def\NIMA{{\em Nucl. Instrum. Methods} A}
\def\NPB{{\em Nucl. Phys.} B}

\def\PRD{{\em Phys. Rev.} D}

\def\IEEE{\em IEEE Trans. Nucl. Sci.}

\def\ra{\rightarrow}

\def\ab{\bar{\alpha}}
\def\be{\begin{equation}}
\def\ee{\end{equation}}
\def\bea{\begin{eqnarray}}
\def\eea{\end{eqnarray}}
\def\sintw{\sin^2(\theta_W^{eff})}
\def\ab{A_b}
\def\rb{R_b}
\def\zbb{Zb\overline{b}}

\def\alr        {A_{LR}}
\def\alrfb	{A_{LR}^{FB}}

\begin{document}
\thispagestyle{empty}
\renewcommand{\thefootnote}{\fnsymbol{footnote}}
\rightline{SLAC-PUB-8143}
\rightline{May 1999}
\begin{center}
{\bf
ELECTROWEAK RESULTS FROM THE SLD EXPERIMENT\footnote{
    Work supported by Department of Energy contract
     DE-AC03-76SF00515}}
\\
\vspace{4mm}
{\bf M. Woods}
\\
\vspace{4mm}
{\small Stanford Linear Accelerator Center,\\
Stanford University, Stanford, CA 94309\\}
\vspace{3mm}
{\small Representing\\}
\vspace{3mm}
{\bf The SLD Collaboration\\}
\end{center}

\begin{center}
\vspace{5mm}
\begin{minipage}{130 mm}
\small
{\bf Abstract.}
We present an overview of the electroweak physics program of the SLD experiment
at the Stanford Linear Accelerator Center (SLAC).  A data sample of 550K 
$Z^0$ decays has been collected.  This experiment utilizes a highly polarized 
electron beam, a small interaction volume, and a very precise pixel vertex 
detector.  It is the first experiment at a linear electron collider. We 
present a preliminary result for the weak mixing angle, $\sintw=0.23110 \pm
0.00029$.  We also present a preliminary result for the parity violating 
parameter, $\ab = 0.898 \pm 0.029$.  These measurements are used
to test for physics beyond the Standard Model.
\end{minipage}
\end{center}
\centerline{{\it Presented at the Lake Louise Winter Institute:  Electroweak
Physics}}
\centerline{{\it Lake Louise, Alberta, Canada}}
\centerline{{\it February 14-20, 1999}}

\pagebreak

\normalsize
\pagestyle{plain}
\section{Electroweak Physics program of the SLD Experiment}

The SLD Experiment began its physics program at the SLAC Linear Collider (SLC)
in 1992, and has accumulated a total data sample of approximately 550K 
hadronic $Z^0$ decays between 1992 and 1998.  
This data sample is a factor 30
smaller than the $Z^0$ sample available from the combined data of the
4 LEP experiments, ALEPH, DELPHI, L3 and OPAL.  Yet the SLD physics results
in many areas are competitive with the combined LEP result, and for some
measurements SLD has the world's most precise results.    

There are 3 features that distinguish the SLD experiment at the SLC:  
a small, stable interaction volume;
a precision vertex detector; and a highly polarized electron beam.  SLD
is the first experiment at an electron linear collider. 
The collision volume is small and stable, 
measuring 1.5 microns by 0.7 microns in the transverse dimensions by  
700 microns longitudinally.  

These key features for the SLD experiment result in the world's best 
measurement of the weak mixing angle,  a precise direct measurement of  
parity violation at the $\zbb$ vertex, $\ab$, and a good measurement of 
the $\zbb$ coupling strength, $\rb$.  The weak mixing angle 
measurement 
provides an excellent means to search for new physics that may enter through
oblique (or loop) corrections, while the $\ab$ and $\rb$ measurements are 
excellent means to search for new physics that may enter through a correction
at the $\zbb$ vertex. 

In its near (analysis) future, SLD is also exploiting its capabilities to 
search for $B_s$ mixing.  The analysis for this is evolving to take full
advantage of the precise vertexing information, and by the time of the summer
1999 conferences SLD should have a measurement of $B_s$ mixing comparable in 
sensitivity with
the combined LEP result.  SLD estimates it should have a reach for $\Delta m_s$
of $12-15 ps^{-1}$, in the region where it is predicted in the SM.  

\vskip 1.5in
\pagebreak
\section{$Z^0$ Coupling Parameters}

At the $Zf\overline{f}$ vertex, the SM gives the vector and axial vector
couplings to be 
$v_f  =  I_f^3-2Q_f \sintw$, and
$a_f  =  I_f^3$,
where $I_f$ is the fermion isospin and $Q_f$ is the fermion charge.
Radiative corrections are significant and are treated as follows.  First,
vacuum polarization and vertex corrections are included in the coupling 
constants, and  an effective weak mixing angle is defined to be
$\sintw \equiv \frac{1}{4}(1-v_e/ a_e )$.  Second,
experimental measurements need to be corrected for initial state radiation
and for $Z-\gamma$ interference to extract the $Z$-pole contribution.

One can define a parity-violating fermion asymmetry parameter, 
$ A_f = \frac{2v_fa_f}{v_f^2+a_f^2} $.
The cross-section for $e^+e^- \ra Z^0 \ra f\overline{f}$ can be expressed by
\begin{equation}
\frac{d\sigma^f}{d\Omega} \propto [v_f^2+a_f^2] \left\{
\begin{array}{c}
(1+\cos^2\theta)(1+PA_e)+ \\
2\cos \theta A_f(P+A_e)  
\end{array}
\right\} \label{eq:xsect}
\end{equation}
where $\theta$ is the angle of the outgoing fermion with respect to 
the incident electron,
and $P$ is the polarization of the electron beam (the positron beam is
assumed to be unpolarized).  We can then define {\it forward, backward}, and
{\it left, right} cross-sections as follows:
$\sigma_F  = \int_{0}^{1} \frac{d\sigma}{d\Omega}d(\cos\theta)$;
$\sigma_B  = \int_{-1}^{0} \frac{d\sigma}{d\Omega}d(\cos\theta)$;
$\sigma_L  = \int_{-1}^{1} \frac{d\sigma_L}{d\Omega}d(\cos\theta)$;
$\sigma_R  = \int_{-1}^{1} \frac{d\sigma_R}{d\Omega}d(\cos\theta)$. 
Here, $\sigma_L$ ($\sigma_R$) is the cross-section for left (right) polarized
electrons colliding with unpolarized positrons.

At the SLC, the availability 
of a highly polarized electron beam allows for direct
determinations of the $A_f$ parameters via measurements of the
{\it left-right forward-backward asymmetry}, $\alrfb$, defined 
by
$$ \alrfb = \frac{(\sigma_F^L-\sigma_F^R)-(\sigma_B^L-\sigma_B^R)}
                  {\sigma_F^L+\sigma_F^R + \sigma_B^L+\sigma_B^R}
          =\frac{3}{4}P_eA_f $$

Additionally, a very precise determination of $A_e$ is achieved from the
measurement of the
{\it left-right asymmetry}, $\alr$, which is defined as
$$\alr = \frac{1}{P_e} \cdot \frac{\sigma_L-\sigma_R}{\sigma_L+\sigma_R} = 
A_e $$
All $Z$ decay modes can be used, and this allows for
a simple analysis with good statistical power for a precise determination of
$\sintw$.
\pagebreak

\section{The SLAC Linear Collider}

\begin{wrapfigure}{r}{6cm}
\epsfig{figure=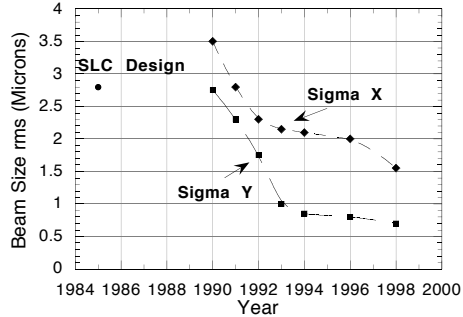,width=6cm}
\caption{Spotsizes at the SLC.  
\label{fig:beamsize}}
\end{wrapfigure}

LEP200 is the last of the large electron storage rings, and a new technology is
needed to push to higher center-of-mass energies.  
Electron linear collider technology provides a means to achieve this, and the
SLC is a successful prototype for this. It has reached
a peak luminosity of $3 \cdot 10^{30} cm^{-2}s^{-1}$, which is within a factor
two of the design luminosity~\cite{raimondi}.  The spotsizes at the 
Interaction Point (IP) are actually 
significantly smaller than design, and Figure~\ref{fig:beamsize} indicates 
how the 
spotsizes have improved with time.  With the small spotsizes, there is
an additional luminosity enhancement from the ``pinch effect'' the two beams
have on each other.  At the higher luminosities achieved in the last
SLC run, the pinch effect enhanced the luminosity by a factor of two.  
\begin{wrapfigure}{r}{4.5cm}
\epsfig{figure=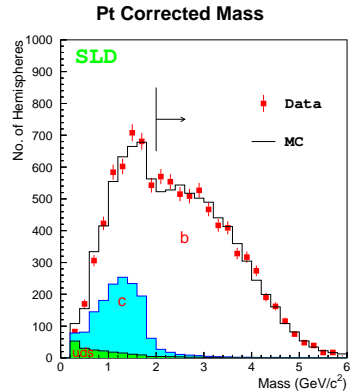,width=4.5cm}
\caption{Vertex mass.  
\label{fig:vtxmass}}
\end{wrapfigure}
The 
luminosity is limited due to the maximum charge achievable in single bunches
because of instabilities in the Damping Rings.

\section{SLD's Vertex Detector}

SLD's vertex detector~\cite{vxd3} consists of 3 layers, with the inner layer 
located at
a radius of 2.7cm.  Angular coverage extends out to $\cos(\theta)=0.90$.  It 
has
307 million pixels, with a single hit resolution of 4.5 microns.  There are
0.4$\%$ radiation lengths per layer. The capability
of SLD's vertex detector is illustrated in Figure~\ref{fig:vtxmass}, which is 
a histogram of the reconstructed jet mass.  With a mass cut of $2.0 GeV/c^2$,
SLD can identify b jets with 50$\%$ efficiency and 98$\%$ purity.

\section{SLD's Compton Polarimeter}

\begin{wrapfigure}{r}{6.5cm}
\epsfig{figure=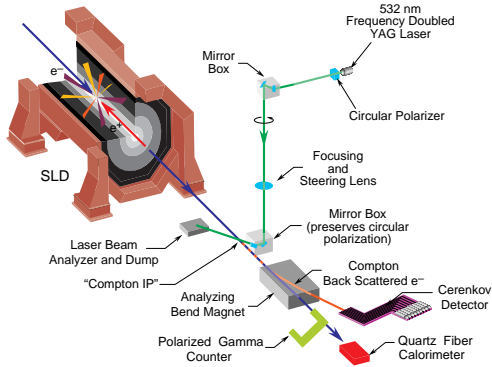,width=6.5cm}
\caption{SLD and the Compton Polarimeter.  
\label{fig:SLD_Compton}}
\end{wrapfigure}

This polarimeter,~\cite{Woods1} shown in Figure~\ref{fig:SLD_Compton}, detects 
both Compton-scattered electrons and Compton-scattered gammas 
from the collision of the longitudinally polarized 45.6 GeV electron
beam~\cite{Woods2} with a circularly polarized photon beam. The photon beam is 
produced from a pulsed Nd:YAG laser with a wavelength of 532 nm.  
After the Compton
Interaction Point (CIP), the electrons and backscattered gammas pass through 
a dipole spectrometer.  A nine-channel threshold Cherenkov detector (CKV) 
measures electrons in the range 17 to 30 GeV.~\cite{monster}
Two detectors, a single-channel
Polarized Gamma Counter (PGC)~\cite{PGC}
and a multi-channel Quartz Fiber Calorimeter (QFC),~\cite{QFC}
measure the counting rates of Compton-scattered gammas.  

Due to beamstrahlung backgrounds produced during luminosity running,
only the CKV detector can make polarization 
measurements during beam collisions.  Hence it is the primary detector and the
most carefully analyzed.  Its systematic error is estimated to be $0.7\%$.
Dedicated electron-only runs are used to compare electron polarization 
measurements between the CKV, PGC and QFC detectors.  The PGC and QFC 
results are consistent with the CKV result at the level of 0.5$\%$.
Typical beam polarizations for the SLD experiment have been in the range
$73-78\%$.

\section{Measurements of $\sintw$, and testing oblique 
corrections}

For the $\alr$ analysis, all {\it Z} decay modes can be used, though in 
practice the leptonic modes are excluded.  They are analyzed separately in
the measurements of $\alrfb$ described
below. The $\alr$ event selection requires at least 4 charged tracks
originating from the IP and greater than 22 GeV energy deposition in the
calorimeter.  Energy flow in the event is required to be 
balanced by requiring the normalized energy vector sum be less than 0.6.  These
criteria have an efficiency of 92$\%$ for hadronic events, with a residual
background of 0.1$\%$.  

SLD's 1998 running yielded 225K
hadronic Z decays, with $N_L=124,404$ produced from the left-polarized beam and
$N_R=100,558$ produced from the right-polarized beam.  For the measured beam
polarization of 73.1$\%$, this yielded $A_{LR}^{meas}=0.1450 \pm 0.0030 
(stat)$.  Correcting for initial state radiation and $Z-\gamma$ 
interference effects, gives
$A_{LR}^0=0.1487 \pm 0.0031(stat) \pm 0.0017(syst)$.  The systematic error 
includes a contribution of 0.0015 from uncertainties in the polarization scale 
and 0.0007 from uncertainties in the energy scale. This result determines the
weak mixing angle to be $ \sintw = 0.23130 \pm 0.00039 \pm 0.00022$.
Combining all of SLD's $\alr$ results from 1992-98, gives $\sintw =
0.23101 \pm 0.00031$.  

\begin{wrapfigure}{r}{6cm}
\epsfig{figure=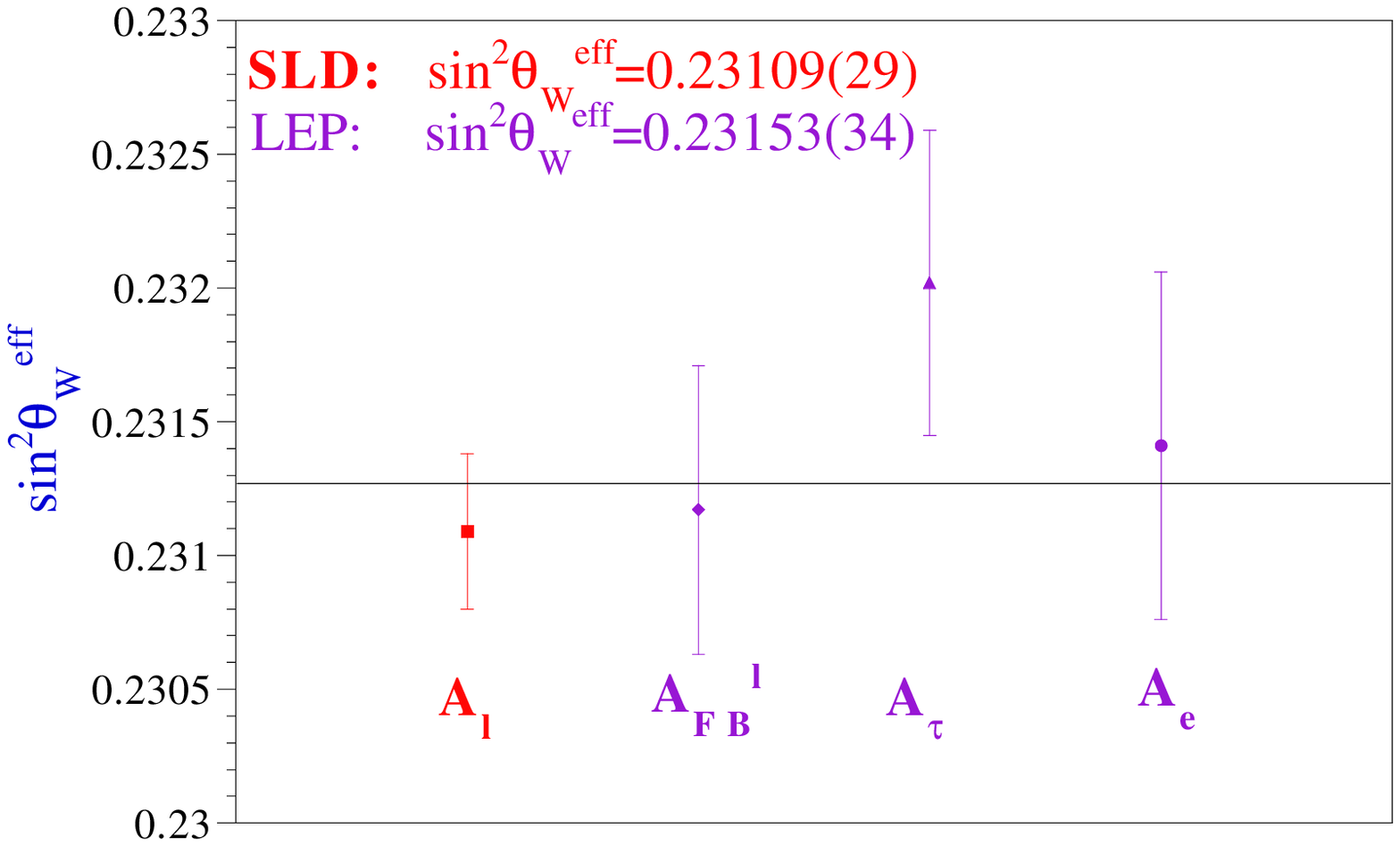,width=6cm}
\caption{Weak Mixing angle measurements.  
\label{fig:aleptons_tech}}
%
\epsfig{figure=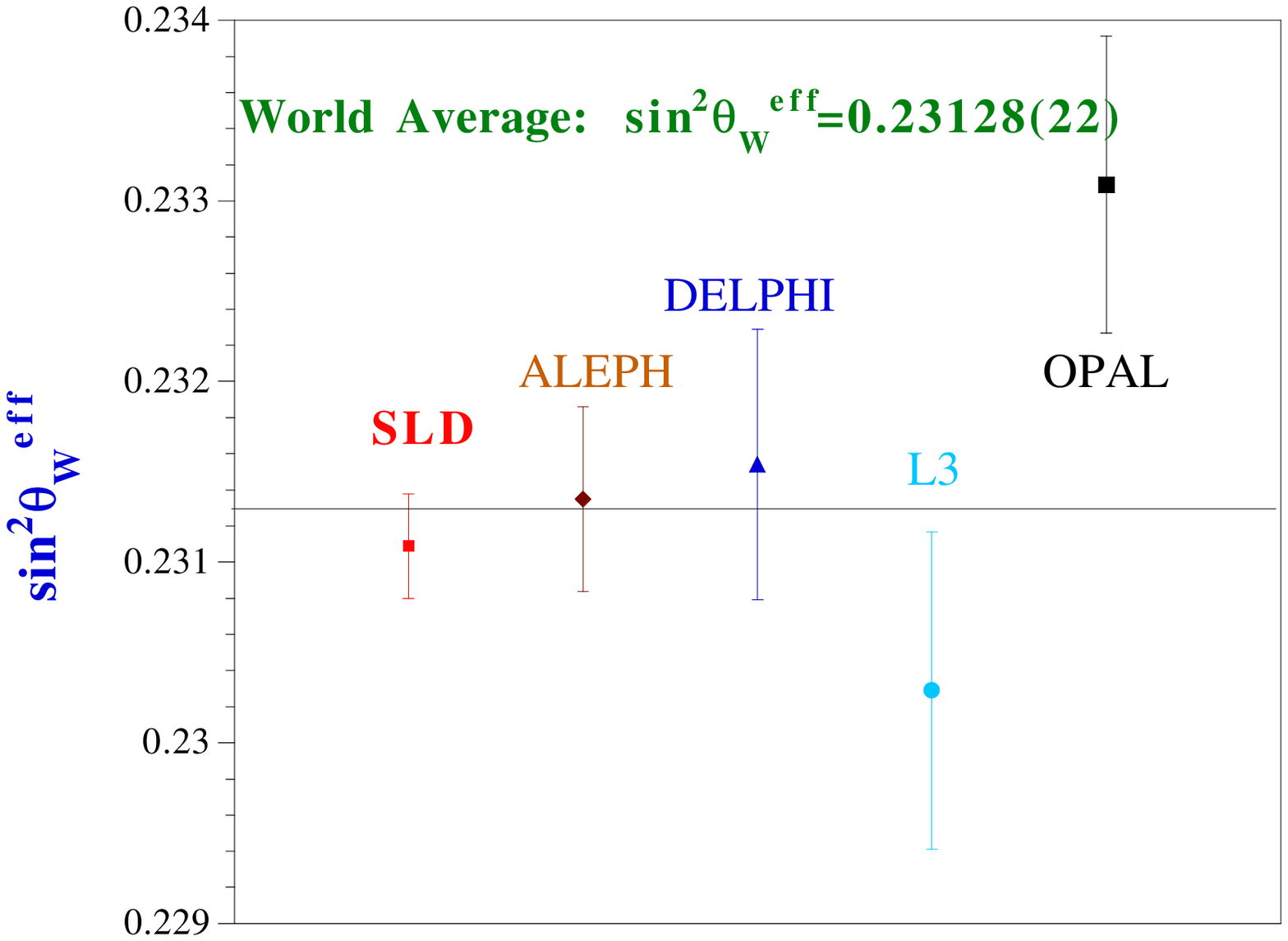,width=6cm}
\caption{Weak Mixing angle measurements.  
\label{fig:aleptons_expt}}
\end{wrapfigure}

For the $\alrfb$ analysis, 
leptonic Z decay events are selected as follows.  The number of charged
tracks must be between 2 and 8.
One hemisphere must have a charge of -1
and the other hemisphere a charge of +1.  The polar angle is required to have
$\cos (\theta) < 0.8$.  For {\it ee} final state events, the 
only additional requirement is a deposition of greater than 45 GeV in the 
calorimeter.  The $\mu \mu$ final state events must reconstruct with a large
invariant mass and have less than 10 GeV per track deposited in the 
calorimeter.  The $\tau \tau$ final state events must reconstruct with an
invariant mass less than 70 GeV, and deposit less than 27.5 GeV per track in
the calorimeter.  One stiff track is required ($>3$ GeV), the acollinearity
angle must be greater than $160^{\circ}$ and the invariant mass in each 
hemisphere must be less than 1.8 GeV.  Event selection efficiencies are
87.3$\%$ for $ee$, 85.5$\%$ for $\mu \mu$, and 78.1$\%$ for $\tau \tau$.
Backgrounds are estimated to be 1.2$\%$ for $ee$ (predominantly $\tau \tau$),
0.2$\%$ for $\mu \mu$ (predominantly $\tau \tau$), and 5.2$\%$ for
$\tau \tau$ (predominantly $\mu \mu$ and $2\gamma$).

We use Equation~\ref{eq:xsect} in a maximum likelihood analysis 
(which also allows for photon exchange and for $Z-\gamma$ interference) 
to determine
$A_e$, $A_\mu$ and $A_\tau$.  The results are  
$    A_e = 0.1504 \pm 0.0072$, 
$  A_\mu = 0.120 \pm 0.019$, and
$ A_\tau = 0.142 \pm 0.019$.  These results are consistent with 
universality and can be combined, giving
$ A_{e,\mu,\tau} = 0.1459 \pm 0.0063$.  This 
determines the weak mixing angle to be $\sintw = 0.2317 \pm 0.0008$.

Combining the $\alr$ measurements that use hadronic final states and the
$\alrfb$ measurements that use leptonic final states,
we determine the weak mixing angle to be 
$\sintw = 0.23110 \pm 0.00029$.
This is a preliminary result.  

A comparison of SLD's result 
with leptonic asymmetry measurements at LEP~\cite{lepew} is given in
Figure~\ref{fig:aleptons_tech}.  These results are compared by technique, 
where $A_l$ is SLD's
combined result from $\alr$ and $\alrfb(leptons)$ described above; $A_{FB}^l$
is the LEP result using the {\it forward-backward} asymmetry with leptonic
final states; $A_\tau$ and $A_e$ are the LEP results from analyzing the 
$\tau$ polarization for the $\tau\tau$ final state.  We do not include in
this comparison the LEP results using hadronic final states.  These results
are discussed below, when we examine SLD's $A_b$ measurement and tests of
{\it b} vertex corrections.
The SLD and LEP data in Figure~\ref{fig:aleptons_tech} are replotted in 
Figure~\ref{fig:aleptons_expt} by experiment rather 
than by technique.  The data are consistent and can be combined to give a world
average $\sintw = 0.23128 \pm 0.00022$.  

\begin{wrapfigure}{r}{6cm}
\epsfig{figure=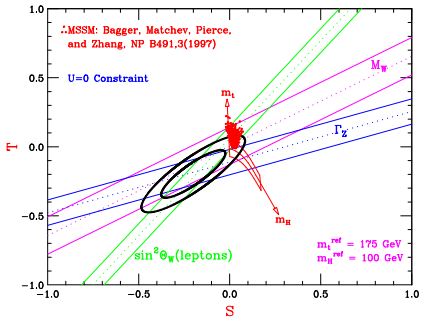,width=6cm}
\caption{Testing oblique corrections.  
\label{fig:stplot}}
\end{wrapfigure}

A convenient framework for analyzing the consistency of the
$\sintw$ measurement with the SM and with other electroweak
measurements is given by the Peskin-Takeuchi parametrization~\cite{Peskin}
for probing extensions
to the SM.  This parametrization assumes that vacuum polarization effects 
dominate
and expresses new physics in terms of the parameters {\it S} and {\it T}, which
are defined in terms of the self-energies of the gauge bosons.  In {\it S-T}
space, a measurement of an electroweak observable corresponds to a band with
a given slope.  Figure~\ref{fig:stplot} shows the {\it S-T} plot for
measurements of the weak mixing angle ($\sintw$), the {\it Z} width 
($\Gamma_Z$),~\cite{lepew} and the {\it W} mass ($M_W$),~\cite{lepew}.
The experimental bands shown correspond to one sigma contours.  
The elliptical
contours are the error ellipses (68$\%$ confidence and 95$\%$ confidence) for
a combined fit to the data. 
The SM allowed region is the small parallelogram, with arrows indicating the
dependence on $m_t$ and $m_H$.
The Higgs mass is allowed to vary from 100 GeV to 1000 GeV and $m_t$
from 165 GeV to 185 GeV.  
The measurements are in reasonable agreement with
the SM and favour a light Higgs mass.  A comparison is also given to a 
prediction for the parameter space of the Minimal Supersymmetric 
Model~\cite{mssm} (region of dots in figure). 
The combined SLD and LEP measurement for the weak
mixing angle gives the narrowest band in {\it S-T} space, and provides
the best test of the SM for oblique corrections.   
Improved measurements of $M_W$ from LEP and FNAL are eagerly awaited to
further constrain and test the SM in this regard.

\section{Measurements of $A_b$, and \\
testing vertex corrections}

\begin{wrapfigure}{r}{6cm}
\epsfig{figure=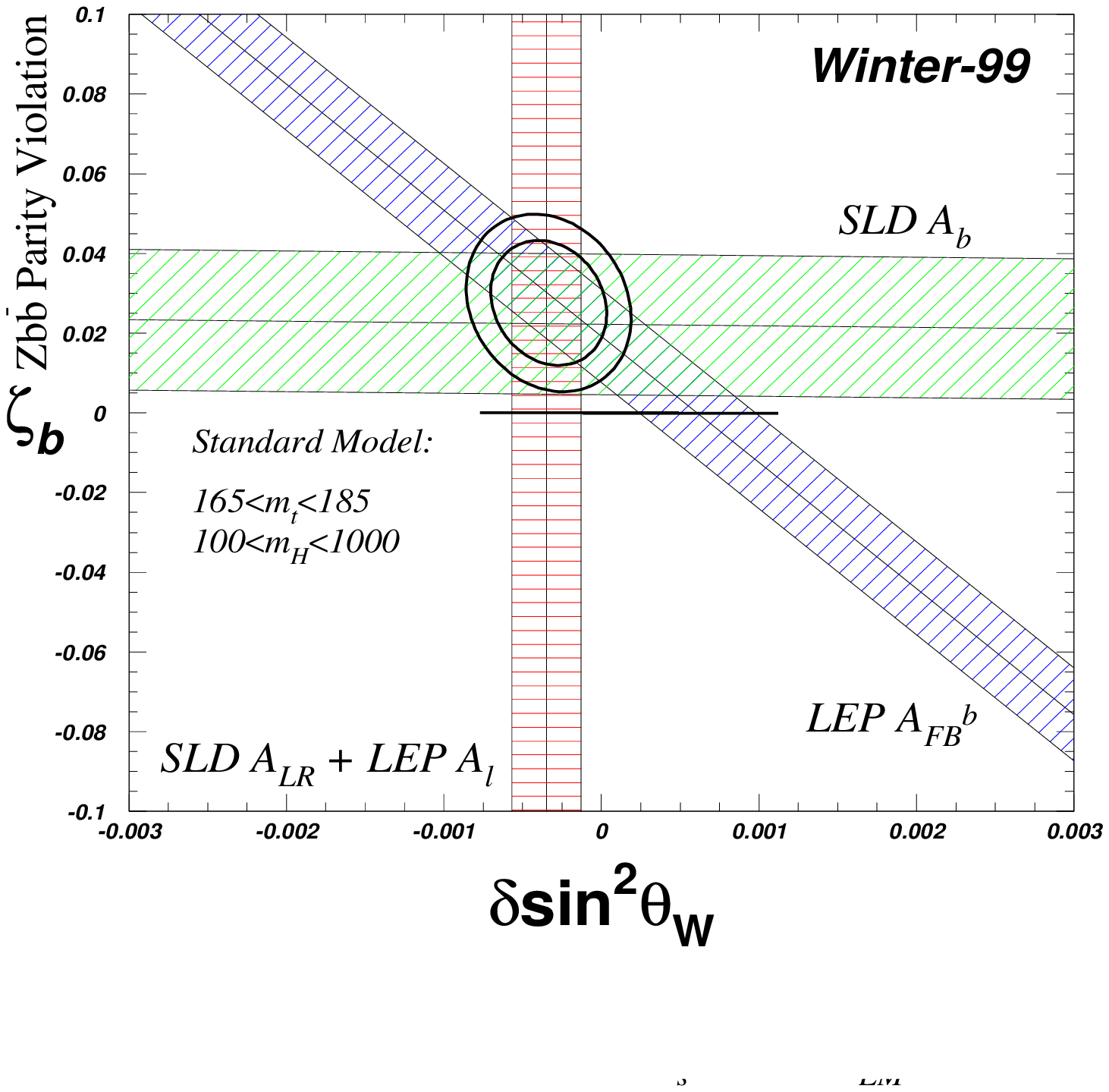,width=6cm}
\caption{Testing vertex corrections.  
\label{fig:tgrplot}}
\end{wrapfigure}

The measurement technique for determining $A_b$ is similar to that for
determining $A_e$, $A_\mu$ and $A_\tau$.  
For this analysis, the capabilities of SLD's vertex detector is
critical and good use is also made of SLD's particle identification system to
identify kaons with the Cherenkov Ring Imaging Detector (CRID).  Three 
different analyses are employed with different techniques for determining the 
{\it b}-quark charge.  The {\bf Jet Charge} analysis uses a momentum-weighted
jet charge to identify the {\it b}-quark charge, and it requires a secondary
vertex mass greater than 2.0 GeV.  The {\bf Kaon Tag} analysis uses the kaon
sign in the cascade decay ($b \ra c \ra s$) to identify the {\it b}-quark 
charge, and it requires a secondary vertex mass greater than 1.8 GeV.
The {\bf Lepton Tag} analysis uses the lepton charge in semileptonic decays
to identify the {\it b}-quark charge; it has no secondary vertex
mass requirement.

The three analyses yield the following results:  $A_b$ (Jet Charge) =
$0.882 \pm 0.020 \pm 0.029$; $A_b$ (Kaon Tag) = 
$0.855 \pm 0.088 \pm 0.102$;
and $A_b$ (Lepton Tag) = $0.924 \pm 0.032 \pm 0.026$.  These results
can be combined, giving $A_b = 0.898 \pm 0.029$.  This is a preliminary result.

Similar to the {\it S-T} analysis for testing oblique corrections, one can
utilize an extended parameter space for testing vertex corrections.
This is done in Figure~\ref{fig:tgrplot}, where we plot the deviation in 
$A_b$ from the SM prediction versus the deviation in $\sintw$ from the SM
prediction.  The three bands plotted are SLD's $A_b$ measurement,
the combined leptonic $\sintw$ measurement from SLD and LEP,
and LEP's {\it forward-backward b} asymmetry.  The elliptical
contours are the error ellipses (68$\%$ confidence and 95$\%$ confidence) for
a combined fit to the data.  The horizontal line is the SM prediction.  We 
note that the data are in excellent agreement, but differ from the SM 
prediction by 2.6$\sigma$.  Unfortunately, there will be no new data 
to indicate
whether this deviation results from a statistical fluctuation, a problem in 
the {\it b} physics analysis, or new physics.  We also note the
discrepancy of where the SLD-LEP $\sintw$ and the LEP $A_{FB}^b$
measurement bands intersect the SM line.  This reflects their 2.2$\sigma$
discrepancy in determining $\sintw$ within the SM framework. 

\section{Conclusions}

The SLD experiment has been the first experiment at an electron linear 
collider.  The viability of a linear collider has been demonstrated and this
technology is now being proposed for future $e^+e^-$ colliders with 
center-of-mass energies up to 1 TeV.  The SLD has made many important 
contributions to precision electroweak physics.  SLD has made the best
measurement of the weak mixing angle, $\sintw=0.23110\pm0.00029$ {\it
(preliminary)}.  This provides a stringent test of oblique corrections; our 
measurement is consistent with SM predictions and favours a light Higgs mass.
SLD makes the only direct measurement of $A_b$, which we determine to be
$A_b = 0.898\pm0.029$ {\it (preliminary)}.  This measurement, together with
measurements by SLD and LEP of $\sintw$ and LEP's measurement of $A_{FB}^b$,
can be used to test {\it b} vertex corrections.  The data are consistent, 
but indicate a 2.6$\sigma$ discrepancy with the SM prediction.



\section*{References}
\begin{thebibliography}{99}

\bibitem{raimondi} P. Raimondi et al., SLAC-PUB-7955 (1998).
\bibitem{vxd3} K. Abe et al., \Journal{\NIMA} {400}{287}{1997}.
\bibitem{Woods1} M. Woods in {\em SPIN96 Proceedings}, ed. C.W. de Jager et
al. (World Scientific, Singapore, 1997), p.843.
\bibitem{Woods2} M. Woods in {\em SPIN96 Proceedings}, ed. C.W. de Jager et
al. (World Scientific, Singapore, 1997), p.623.
\bibitem{monster} M. Fero et al., SLD-Physics-Note-50 (1996).
\bibitem{PGC} R.C. Field et al., \Journal{\IEEE}{45}{670}{1998}.
\bibitem{QFC} S.C. Berridge et al., in {\em Calorimetry in High Energy Physics
Proceedings}, ed. E. Cheu et al. (World Scientific, Singapore, 1998), p. 170.
\bibitem{lepew} D. Abbaneo et al., CERN-EP-99-015 (1999).
\bibitem{Peskin} M.E. Peskin and T. Takeuchi, \Journal{\PRD}{46}{381}{1992}
\bibitem{mssm} J. Bagger et al., \Journal{\NPB} {491}{3}{1997}.
\bibitem{tgr} T. Takeuchi et al., published in {\em DPF 94 Proceedings}, ed. S.
Seidel (World Scientific, Singapore, 1995), p. 1231.
\end{thebibliography}
\end{document}